\documentclass[oneside,a4paper,12pt,shownumbers,manualsort]{article}
\usepackage{latexsym}
\usepackage{euscript}
\usepackage{epsfig,amsmath,amssymb}

\def\be{\begin{equation}}
\def\ee{\end{equation}}
\def\bea{\begin{eqnarray}}
\def\eea{\end{eqnarray}}

\long\def\symbolfootnote[#1]#2{\begingroup%
\def\thefootnote{\fnsymbol{footnote}}\footnote[#1]{#2}\endgroup} 

 \large\normalsize

\begin{document}
%\draft

\begin{center}

{\Large \bf GUT cosmic strings and inflation}

\vspace*{7mm} {\ Rachel Jeannerot
\symbolfootnote[1]{{E-mail:jeannerot@lorentz.leidenuniv.nl}}}
\vspace*{.25cm}

{\it Instituut-Lorentz for Theoretical Physics,
Niels Bohrweg 2, 2333 CA Leiden, The Netherlands}
\vspace*{.25cm}

\end{center}

\begin{abstract}

We show that GUT cosmic strings generically form after inflation if a
 {\it non-inert} symmetry breaks after inflation; they form
 irrespectively of the inflationary scenario and in both
 supersymmetric and non-supersymmetric theories.

\end{abstract}

In this note, we point out that the formation of field theory cosmic
strings after inflation when Spontaneous Symmetry Breaking (SSB) takes
place is generic. In Ref.\cite{prd} it was conjectured that cosmic
strings always at the end of both standard F-term and D-term inflation
\cite{Fterm,Dterm}.  In Ref.\cite{prd1}, it was shown that cosmic
strings always form in the case of SO(10) GUT with standard hybrid
inflation \cite{hybrid,Fterm}. In Ref.\cite{jrs}, an exhaustive study
of topological defect formation in all possible SSB patterns in all
GUT phenomenologically acceptable was made; it was found that indeed
cosmic strings always form if standard F-term inflation is assumed
\cite{Fterm}; the strings could be topological or embedded. In view of
the revival interest in cosmic (super)-strings \cite{strinfl}, it is
important to clarify and put forward the idea of Ref.\cite{prd}. We
show that the formation of cosmic strings is quite general and does
not rely on the standard hybrid inflationary scenario.  The only
assumption is that inflation solves the monopoles problem and is
followed by a phase transition associated with spontaneous symmetry
breaking.  We do not discuss the nature of the strings which can be
topological \cite{Kibble} or embedded \cite{embedded} and, with non
minimal Higgs content, semi-local
\cite{semilocal}\footnote{Non-topological cosmic strings may remain
stable down to low energy. If the strings form but rapidly decay, they
can still lead to observable effects \cite{lepto}. In some
circumstances, unstable strings may not form \cite{ananot}.}. The
reason why cosmic string form after inflation is a simple argument. In
a nutshell, they form because the group which is broken after
inflation must contain a U(1) factor, and its rank is lowered by (at
least) one unit.

The paper is organized s follows. First, the GUT monopole problem is
reviewed. Inflation as a solution to the monopole problem is then
discussed. It is then shown that if inflation is followed by (at least
one) phase transition during which spontaneous symmetry breaking takes
place (in the visible sector) and no unwanted defect form after
inflation, cosmic strings form.

\paragraph{The monopole problem}

In a cosmological context, when a gauge group $G$ spontaneously breaks
down to a subgroup $H$ of $G$, topological defects form if the vacuum
manifold $G/H$ is non-trivial \cite{Kibble}. Monopoles and domain
walls are cosmologically catastrophic whereas numerical simulations
show that the density of a cosmic string network eventually reaches a
scaling solution, undependently of the initial string density
\cite{ShelVil}. The reason is that cosmic strings can loose energy by
forming loops which rapidly decay via gravitational radiation and
particle emission \cite{ShelVil}.

Undependently of the initial monopole density, monopole-antimonopole
pairs annihilate until the monopole-to-entropy ratio reaches its final
value \cite{preskill,ShelVil}
\begin{equation}
{n_M\over s} \sim {1\over h^6 \beta \sqrt{g_{*s}}} {m \over M_{pl}}
\end{equation}
where $n_M$ is the monopole density and $s$ is the entropy. $h$ is the
monopole magnetic charge which is given by $h=-{4 \pi \over g}$, where
$g$ is the gauge coupling constant. $\beta \sim (1-5) g_*$, where
$g_*$ is the effective number of degrees of freedom and $g_{*s}$ is
the effective number of helicity states for particles with mass $m+p <
T$. Finally, $m$ is the monopole mass and $M_{pl}$ is the Planck
mass. The monopole mass is bounded from below by \cite{Bogomol}
\begin{equation}
m \geq {4 \pi \over e} M_M
\end{equation}
where $M_M$ is the scale at which the monopoles form. For monopoles
forming between the GUT scale $M_{\rm GUT} \simeq 3 \times 10^{16}$ GeV
and the electroweak scale $M_Z \simeq 10^2$ GeV, the monopole density 
\begin{equation}
\Big({n_M\over s}\Big)_{Theo} \in [10^{-12} - 10^{-27}].\label{eq:monotheo}
\end{equation}

Observations imply a much lower density. The less stringent bound
comes from neutron stars observations; the latter can trap monopoles
which can catalyse proton decay and thus increase the star
luminosity. Limits on the luminosity of neutron stars imply a bound on
the monopole flux which translates into a bound on the monopole
density given by \cite{KT}:
\begin{equation}
\Big({n_M \over s}\Big)_{Exp} \leq 10^{-31}. \label{eq:monobound}
\end{equation}

Comparing theoretical predictions Eq.(\ref{eq:monotheo}) with
observations Eq.(\ref{eq:monobound}), we conclude that GUT monopoles
have to be diluted, undependently of the scale at which they
form.

We now turn to the conditions under which monopoles form. Let's
consider a gauge group $G$ which spontaneously breaks down to a
subgroup $H$ of $G$. The breaking can be direct, $G \rightarrow H$, or
via one or more intermediate symmetry subgroups, $G \rightarrow \dots
\rightarrow K \rightarrow \cdots \rightarrow H$. Classification of
topological defects is usually done by using homotopy theory. Domain
walls form when $\pi_0(G/H) \neq I$; they thus form when a discrete
symmetry is broken. Cosmic strings form when $\pi_1(G/H) \neq I$; they
thus form when an abelian symmetry is broken. And monopoles form if
$\pi_2 (G/H) \neq I$. If $G$ is connected (i.e. $\pi_0(G) = I$) and
simply connected (i.e. $\pi_1(G) = I$) then $\pi_2 (G/H) \cong \pi_1
(H)$.  Monopoles which form during a phase transition during which $G
\rightarrow H$ remain stable during the next phase transition during
which $H \rightarrow F$ if $\pi_2 (G/H) \neq I$. Note that if $G$ is
not simply connected, we can always work with its universal covering
group which is simply connected \cite{Kibble}; for example, in the case
of SO(10) which is not simply connected, we work with its universal covering
group, Spin(10), which is simply connected. We are only interested in
the formation of monopoles, and hence we assume that $G$ does not
contain any discrete symmetry, i.e. that it is connected.

Let first $G = G_{GUT}$ a unified gauge group which does not contain a
U(1) factor. It can be simple or semi-simple. Most unified models fit
in this category. For examples, $G_{GUT}$ can be the Georgi-Glashow
model SU(5), SO(10), E(6), the Pati-Salam model $SU(4)_c \times
SU(2)_L \times SU(2)_R$, the trinification $SU(3)_c \times SU(3)_L
\times SU(3)_R$ or $SU(6)$. $\pi_0(G) = I$ and $\pi_1(G) = I$ (if $G$
is not simply connected we consider its universal covering group). $G$
must be broken directly or via intermediate steps down to the
Standard Model gauge group $G_{SM} = SU(3)_c \times SU(2)_L \times U(1)_Y
\equiv H$. Now $\pi_1 (G_{SM}) = Z$ because of its U(1) factor ($\pi_1
(SU(3)) = \pi_1 (SU(2))= I$). Hence $\pi_2 (G_{GUT}/G_{SM}) = Z$ and
therefore monopoles always form at some stage of the SSB breaking
pattern. Furthermore, since $\pi_1(SU(3)_c \times U(1)_Q) = Z$, the
monopoles are topologically stable down to low energy.

Now let's assume that $G_{GUT}$ contains a U(1) factor
which is not {\it inert}. We call an {\it inert} symmetry, a symmetry
which is orthogonal to the Standard Model group. This is for example
the case of flipped SU(5), $\tilde{SU(5)} \times U(1)$. In this case,
$\pi_2 (G_{GUT}/G_{SM}) = I$ and the formation of monopoles can be
avoided \cite{flipped}.

In the final case, we assume that $G$ contains a U(1) factor which is
{\it inert}. So $G$ can be written as $G_{vis} \times U(1)_I$ and
$G_{SM}$ is fully embedded in $G_{vis}$. In that case, one should
consider the breaking of $G_{vis} \rightarrow G_{SM}$ and $U(1)_I
\rightarrow I$ separately; since $\pi_2 (G_{vis}/G_{SM}) = Z$
topological monopoles form.

The conclusion is that as soon as the $U(1)_Y$ symmetry of the
Standard Model is embedded in a non-abelian group which does not
contain a U(1) factor, unwanted monopoles form.  This is the famous
GUT monopole problem \cite{preskill}.

\paragraph{Inflation as a solution to the monopole problem}

A solution to the monopole problem is inflation. - It is often
forgotten that inflation was originally invented for this purpose
\cite{Guth}.- Inflation solves the monopole problem if it takes place
after the phase transition which leads to their
formation. Alternatively the phase transition can take place during
inflation, but in the case of GUT scale monopole it must be at least
some $19$ e-folds before the end of inflation \cite{therm}. Another
solution to the monopole problem would be to have a Langacker-Pi type
phase transition \cite{Lpi}. We focus on inflation.

\paragraph{GUT cosmic strings are produced after inflation}
We now show that cosmic strings are always produced after inflation if the
following assumptions are satisfied :

\begin{enumerate}

\item The Standard Model is embedded in a semi-simple grand unified
 gauge group $G_{GUT}$

\item Inflation solves the monopole problem

\item Spontaneous Symmetry Breaking (SSB) in the visible sector takes
place after inflation\footnote{This is always true in the case of
standard hybrid inflation \cite{Fterm}.} (apart from Standard Model breaking).

\end{enumerate}

The grand unified gauge group $G_{GUT}$ must break down to
$G_{SM}$. The breaking can be direct, $G_{GUT} \rightarrow G_{SM}$, or
via intermediate symmetry subgroups, $G_{GUT} \rightarrow
... \rightarrow G_{SM}$.  Whereas direct breaking is allowed in the
supersymmetric case, at least one intermediate symmetry group is
needed for unification of the coupling constants if there is no
supersymmetry. Particle physics can hardly constrain the spontaneous
symmetry breaking patterns\footnote{Accelerator experiments can
constrain the intermediate groups if they break at very low
scale. Some intermediate breakings are excluded because they lead to
rapid proton decay. In some cases, neutrinos constrains can also be
used. From theoretical side, patterns with large number of
intermediate steps are disfavored.}.  Here, we use cosmology
\cite{prd,prd1}. Since SSB is assumed to take place after inflation,
$G_{GUT}$ cannot break directly down to $G_{SM}$, otherwise monopoles
would form after inflation. Therefore there must be at least one
intermediate symmetry group, let's call it $K$, such that $G_{GUT}$
breaks down to $K$ before inflation and the monopoles
form \footnote{The breaking must happen at least 19-efolds
before the end of inflation.}, and no unwanted defect form when $K$
breaks down to $G_{SM}$ after inflation
\footnote{The breaking of $K$ is a dynamical problem which usually
depends upon the coupling (direct or indirect) between the inflaton
and the GUT Higgs field(s), but not necessarily on the reheating
temperature. For example in the case of hybrid inflation, if the Higgs
field which triggers the end of inflation is the Higgs field used to
break $K$, the breaking of $K$ occurs undependently of the value of
the reheating temperature.}. $G_{GUT} \supset K \supset G_{SM}$. There
can be intermediate breaking(s) between $G_{GUT}$ and $K$ and between
$K$ and $G_{SM}$:
\begin{equation}
 G_{GUT} \overset{\rm Monopoles}{\rightarrow \dots \rightarrow} K
 \overset{\rm Inflation, \,\, \, No \,\, unwanted \,\,
 defect}{\rightarrow \dots \rightarrow} G_{SM}.
\end{equation}
For example, $K$ cannot contain any discrete symmetry otherwise
 unwanted domain walls would form (unless this discrete symmetry
 remains unbroken down to low energy).

Since monopoles form when $G_{GUT}$ breaks down to $K$, the latter
 contains at least one U(1) factor, which is not {\it inert},
 otherwise monopole would again form when $K$ breaks down to
 $G_{SM}$. We call it $U(1)_X$. $K$ can contain more abelian factor(s)
 ({\it inert} or not); $\pi_1(K) = Z^n$ if $K$ contains $n$ U(1)
 factors. $K$ is thus of the form
\begin{equation}
K = J \times U(1)_X \times \underbrace{U(1) \times \dots \times U(1)}_{n-1} 
\end{equation}
where $J$ a semi-simple group and $n \geq 1$. 

We now consider the various possibilities for $n$ and $J$ and the consequences
for defect formation.

\begin{enumerate}

\item [Case 1:] $n>1$

In this case, the first homotopy group $\pi_1(K/G_{SM}) = Z^{n-1} \neq
I$ and therefore topological cosmic strings always form. They form
regardless of the embedding of flavor, color and electric charge in
$K$. The rank of $K$ is strictly greater than the rank of the Standard
Model gauge group, i.e. rank$(K) \geq 5$, and rank$(G_{GUT}) \geq 5$.

\item [Case 2:] $n=1$

In this case, the intermediate symmetry group $K = J \times U(1)_X$,
 $J \supset SU(3)_c \times SU(2)_L$ and $G_{SM}$ is fully embedded in
 $J \times U(1)_X$.  There are then two possibilities. 

\begin{enumerate}

\item  $U(1)_X = U(1)_Y$

$SU(3) \times SU(2)$ can only be embedded in a group with rank $\geq
4$. Thus the rank of $J$ is $\geq 4$ and the rank of $K$ is $\geq
5$. There are a priori five possibilities for $J$.

\begin{enumerate}

\item $J=SU(3)_c \times SU(2)_L \times F$ with $F$ simple or semi
simple. $F$ is {\it inert}. rank$(G_{GUT}) \geq 6$. Cosmic strings do
not form. There are transient strings connecting monopole-antimonopole
pairs if F breaks down to the identity in more than one step; for ex
$F \rightarrow U(1) \rightarrow I$.

\end{enumerate}

The remaining ones are
\begin{enumerate}
\addtocounter{enumiii}{1}

\item $J = SU(3)_c \times F$ with $SU(2)_L \supset F$ 

\item $J = F \times SU(2)_L$ with $SU(3)_c \supset F$ 

\item $J = F \times P$ with $SU(3)_c \supset F$ and $SU(2)_L \supset
P$ 

\item $J = SU(3)_c \times SU(2)_L \supset F$ 

\end{enumerate}
with $F$ and $P$ simple or semi simple (do not contain a U(1) factor).
These cases which, if at all allowed, correspond to very
non-standard embeddings of color, flavor and electric charge and very
non-standard GUTs. There is no string.

\item $U(1)_X \neq U(1)_Y$

In this case, the hypercharge is a linear combination of $X$ and of
(at least) one diagonal generator of $J$. The orthogonal generator is
broken and embedded strings form \cite{embedded}. 
%This case corresponds to standard GUTs and standard embedding. 

\end{enumerate}

\end{enumerate}
\paragraph{From embedded to topological strings}

Discrete $Z_N$ symmetries are commonly left unbroken in realistic SSB;
this depends upon the Higgs representation which is used to do the
breaking\footnote{The most common example is the case of GUT gauge
groups which contain gauged $B\!-\!L$ and predict neutrinos masses via
the see-saw mechanism. If $B\!-\!L$ if broken with Higgs fields in
``safe'' representations of $G_{GUT}$ \cite{martin}, a discrete $Z_2$
symmetry (which plays the role of matter parity) subgroup of $B-L$ is
left unbroken, and the proton decays at an acceptable rate. If it is
broken, R-parity has to be imposed by hand.}. If a discrete $Z_N$
symmetry is left unbroken after the breaking of $K$, in cases
2.(b) topological strings form instead of embedded
strings, and in cases 2.(a), $Z_N$ strings form. 

\vspace{.2cm}

In conclusion, we have considered generic GUT models where SSB takes
place after inflation. There is no assumption made about the
inflationary scenario.  The model can be supersymmetric or
not. Inflation solves the monopole problem if there is at least one
intermediate symmetry group between $G_{GUT}$ and $G_{SM}$.  We have
shown that the rank of the intermediate group must be greater or equal
to $5$ and is lowered at the end of inflation; hence the rank of
$G_{GUT}$ must also be greater or equal to $5$. We have also shown
that there must be at least one intermediate symmetry group of the
form $J \times \underbrace{U(1) \times \dots \times U(1)}_{n}$ where
$J$ is a semi-simple group and $n \geq 1$. We found that cosmic
strings always form after inflation if the symmetry which breaks after
inflation is not {\it inert}. The strings could be topological or
embedded. When a discrete symmetry is left unbroken after the breaking
of the intermediate symmetry group, the strings are always
topological.  In some particular very non standard embeddings of color
and flavor, which do not apply to any known GUT, cosmic strings do not
form. If the strings are stable, they contribute to primordial
fluctuations \cite{CSCMB}. However their contribution may be too low
for detection via temperature anisotropies of the CMB. For example, in
the case of hybrid inflation, scalar perturbations are dominated by
scalar perturbations from inflation for most of the parameter space
\cite{cmb}. On the other hand, cosmic strings may well be detected via
B-type polarization of the CMB \cite{Seljak}. In models with hybrid
inflation for example, tensor perturbations are negligible and vector
and tensor perturbations are dominated by perturbations from cosmic
strings. If the strings decay before nucleosynthesis, they will not be
detected via CMB experiments. They may however have interesting
cosmological consequences \cite{infl}.

\section*{Acknowledgments}

The author would like to thank A. Achucarro and J. Polchinski for
discussions. She would also like to thank the Netherlands Organization
for Scientific Research [NWO] for financial support.


\begin{thebibliography}{99}
\bibitem{ananot} 
This was pointed out to us by Ana Achucarro.
\end{thebibliography}
\end{document}